# Formation of Laves Phases in Repulsive and Attractive Hard Sphere Suspensions

N. Schaertl[1,2], T. Palberg[3], E. Bartsch[1,2]

[1] Institut für Makromolekulare Chemie, Albert-Ludwigs-Universität Freiburg, 79104 Freiburg, Germany

[2] Institut für Physikalische Chemie, Albert-Ludwigs-Universität Freiburg, 79104 Freiburg, Germany

[3] Institut für Physik, Johannes Gutenberg Universität Mainz, 55128 Mainz, Germany

**Colloidal Laves phases (LPs) of $MgCu_2$ type are promising precursors for diamond structure photonic materials. They have been predicted for hard sphere binary mixtures, but not yet observed. We here report a time resolved static light scattering study on their formation in a binary mixture of buoyant experimental hard sphere approximants (size ratio $\Gamma=0.77$, molar fraction of small spheres $x_S = 0.76$) for volume fractions between melting and the glass transition. In line with theoretical expectation, all samples form LPs of $MgZn_2$ structure on the time scale of weeks to months. $MgNi_2$ structures are absent, $MgCu_2$ structures and randomly stacked LPs prevail at elevated volume fraction. The addition of small amounts of non-adsorbing polymer switches the interaction to depletion attractive and results in significantly accelerated crystallization kinetics and improved crystal quality.**

Corresponding author: E. Bartsch **eckhard.bartsch@physchem.uni-freiburg.de**

Laves phases (LPs) have first been discovered in binary metallic systems and since have been a field of continued interest.[1, 2, 3, 4, 5] Their colloidal analogues are expected to be promising precursor candidates for photonic materials with diamond or pyrochlore structure.[6] A prominent model for colloidal particles are hard sheres (HS). These show an entropy driven first order freezing transition at sufficiently large volume fraction, $\Phi = n(3/4\pi R^3)$ (Here n denotes the particle number density and R the HS radius).[7] Also binary mixtures of HS have been extensively studied. While alloy formation at large total volume fraction can be rationalized solely by packing arguments, non-close-packed structures like LPs or $AB_{13}$ are stabilized by contributions from free volume or phononic entropy.[8, 9] This also applies for the LPs with $MgZn_2$, $MgNi_2$ or $MgCu_2$ structure which differ only by the particular stacking of the same four-layered structural units. These LPs were recently predicted to be the structures of lowest free energy in binary HS mixtures in a narrow range of size ratios $0.76 \leq \Gamma = R_S/R_L \leq 0.84$.[6,10] Other theoretical studies found purely eutectic behavior for the size ratio of $\Gamma = 0.816$, which yields the maximum packing fraction for the $LS_2$ LPs.[11,12,13]

Interestingly, despite numerous experimental studies on binary colloidal mixtures,[14,15,16,17,18,19] only few observations of these elusive structures have been reported so far. LPs were found in light microscopic studies of binary charged sphere mixtures,[20,21] in scattering studies on moderately concentrated charged sphere mixtures of very large polydispersity,[22] in mixtures of latex colloids from different materials,[23] in natural Silicate Opals[24], and in electron microscopy studies on dried, ligand-stabilized nano-particle suspensions.[25,26,27,28] None of these systems is a simple hard sphere (HS) binary mixture. These observations therefore indicate some additional subtle influence of the type and range of interactions, as is also discussed in connection with intermetallic LPs.[4,5] Moreover, they present a challenge to experimentalists to conduct studies on very close hard sphere approximants.[29]

In the present study, we use extremely well buoyancy matched spherical particles with excluded volume interactions only, which are characterized by a very steep repulsive pair interaction of $U(r) \propto (r)^{-n}$, with n = (40-46). We study a binary mixture with a size ratio close to $R_S/R_L = 0.77$ at a molar fraction $x_S = N_S/(N_S+N_L) = 0.76$, where S denotes the small and L denotes the large spheres and $N_{S,L}$ are their respective numbers. We further cover a large range of volume fractions, $\Phi$ between melting and the glass transition. These parameters cover the coexistence of LPs with several other phases and moreover are close to those of the largest extension of the LP – Fluid (F) coexistence region in which we expect the LP forma-tion to proceed in the absence of all other solidification processes.[10] While our experiments required some patience, we were rewarded by the observation of spontaneous formation of LP crystals via homogeneous nucleation from the shear homogenized colloidal melt for all volume fractions investigated.

We further repeated these experiments for the sample at volume fraction $\Phi = 0.578$ under addition of different amounts of non-adsorbing polymer, thereby introducing an attractive component into the potential of mean force. Such depletion attractions have been extensively studied in experiment and theory, but neither LPs have been observed nor predictions of the phase diagram of attractive binary mixtures are avilable. In a previous study on a eutectic binary mixture of size ratio $R_S/R_L = 0.74$, we found that its eutectic phase behavior was retained upon addition of polymer.[30,31] Moreover, after suppressing the intervening glass transition the complete phase diagram, including the coexisting phases of pure L and S became accessible. The presently investigated region of LP formation is adjacent to the fluid phase at comparably low volume fraction, such that no glass transition should suppress their observation. However, also in the present study, we find that phase behavior, crystallization kinetics and the resulting crystal quality can be manipulated via the addition of polymer. We therefore believe that our findings will stimulate further theoretical investigations on phase transition kinetics and advance the realization of pyrochlore or diamond structure based photonic materials.

Given the concerns about correct preparation and characterization,[29,32] the most important prerequisites for our study on the existence and formation kinetics of HS-LPs are very carefully buoyancy matched particles of HS-like interactions and a precisely chosen size ratio. We used 1:50 cross-linked (1 crosslink per 50 monomer units) polystyrene (PS) micro-gel spheres which were synthesized by emulsion polymerization, cleaned, dried and re-suspended in 2-ethylnaphthalene (2EN), as described previously.[33] As compared to other HS approximants,[29] this system shows an excellent simultaneous match of both its refractive index and its mass density to those of the solvent. At T = 20°C, the bulk refractive indices at $\lambda = 633$ nm and mass densities were determined to be $n_{633}(2EN) = 1.594$, $n_{633}(PS, 1:50) = 1.602$, $\rho(2EN) = 0.992$ g cm$^{-3}$ and $\rho(PS) = 1.05$ g cm$^{-3}$, respectively. These already small differences are further reduced as the spheres swell to reach their equilibrium radii. This way HS systems under near µ-g conditions have been realized in which crystallization was observed for volume fractions up to $\Phi \leq 0.585$.[34]. We therefore expect no influence of sedimentation nor jamming for the present study.

The "hardness" of micro-gel particles can be obtained from rheological measurements[35] interpreted in terms of an inverse power-law pair interaction $U(r) \propto (r)^{-n}$, where n is the hardness exponent. Computer simulations show that for $n \geq 18$, the resulting potentials are sufficiently steep to approximate the particles as HS.[36] Sub-micron sized PMMA particles exhibit values around n=150.[37] In a comprehensive study on particles with different sizes and different degrees of cross-linking, we recently studied the dependence of n on synthesis conditions.[33] We found that n shows a non-linear decrease with cross-linking density with a considerable spread for different particle sizes, but also a clear inverse power law dependence, $n \propto Q^{-1}$, on the volume swelling ratio $Q = R_{PD}^3/R_u^3$, with the radius in the un-swollen state, $R_u$, measured by transmission electron microscopy. These investigations confirmed the previously reported value for 1:50 cross-linked particles studies of $n = 40\pm10$.[30]

FIG. 1 shows the phase diagram of a binary HS mixture of size ratio Γ 0.76 as reproduced from the simulations of Hynninen et al..[10] It contains three stable phases, one component crystals of S or L particles and the Laves phase, LP. In FIG 1a we show the reduced pressure – composition plane. The LPs have a molar fraction $x_S = 0.67$. The pocket of F-LP coexistence adjacent to the stable fluid (F) phase extends from $x_S \approx 0.73$ out to $x_S \approx 0.9$ with a considerable extension also in p-direction. For all other investigated size ratios investigated in ref. 10, it was found to be significantly smaller. FIG 1b shows the coexistence lines of FIG. 1a widened to coexistence regions in the $\phi_S$ - $\phi_L$ plane with $\phi_{S,L}$ denoting the volume fraction of each component.

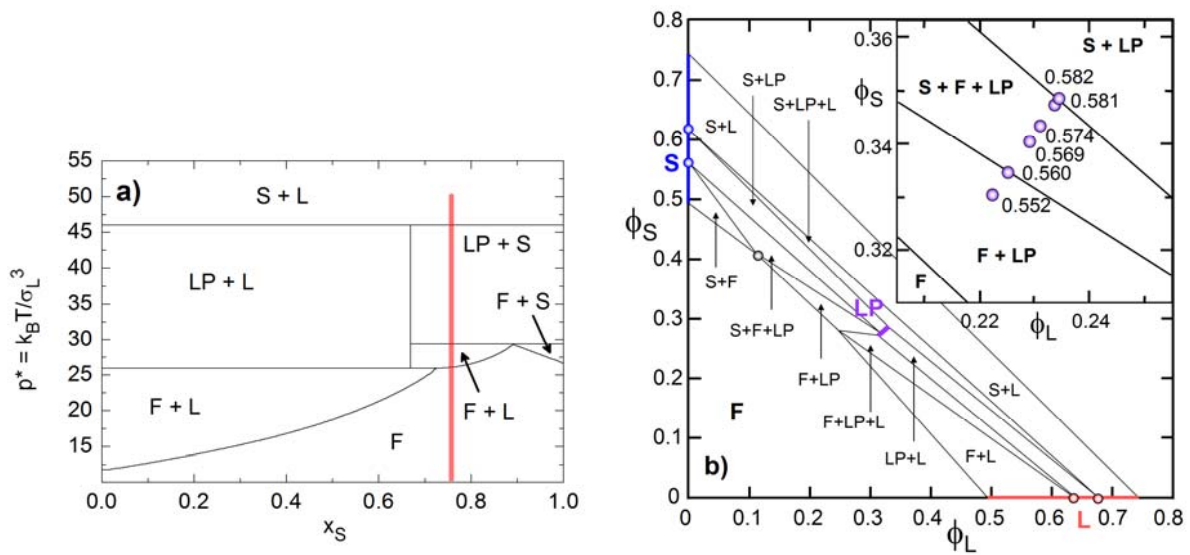

FIG. 1: a) Phase diagram of binary HS mixtures of size ratio Γ = 0.76 in the reduced pressure – molar fraction plane. The vertical line indicates the molar fraction of our experimentally studied systems of size ratio Γ = 0.77:

$x_S$ = 0.76 (number ratio Z=3.2; volume fraction ratio $\phi_S/\phi_L$=1.5) and its estimated relative extension under variation of overall volume fraction and polymer concentration. b) $\phi_S$ - $\phi_L$ representation of the same phase diagram. Inset: enlargement with violet dots representing the location of the R-samples with total volume fractions, $\Phi = \phi_S + \phi_L$, as indicated. (Redrawn from ref. 10 with kind permission of Springer Verlag).

Particles suited for a reliable comparison were pre-selected by their hydrodynamic size ratio, $\Gamma_h = R_{h,S}/R_{h,L}$. This quantity is conveniently determined from dynamic light scattering and may serve as a first orientation. However, further characterization is warranted since $\Gamma_h$ may be significantly affected by the unknown surface roughness of each particle species. Much better estimates of $\Gamma$ can be obtained from phase diagram mapping[38, 39], yielding $\Phi_{PD}$, $R_{PD}$ and $Q_{PD}$ or the location of the first Bragg peak for crystals at coexistence, yielding $\Phi_{BR}$ and $R_{BR}$.[40] However, these two approaches suffer from polydispersity effects. In the first case, one maps the unswollen volume fraction onto the well known freezing volume fraction of monodisperse hard spheres, $\Phi_{F,HS}$ = 0.494. This completely neglects polydispersity. Some improvement can be gained by using freezing points calculated for polydisperse systems.[41, 42, 43, 44] However, an exact determination of the polydispersity $\sigma = \sqrt{\langle R^2 \rangle - \langle R \rangle^2}/\langle R \rangle$ is difficult since unavoidable multiple scattering effects lead to a systematic overestimation of $\sigma$.[45] Moreover, incomplete knowledge of the shape of the particle size distribution hinders a precise comparison to the predictions. In the second case, fractionation effects may lead to a deviation of the average size in the fractionated crystals from that of the original melt.[46,47] Thus, even if the were well known for the swollen state from form factor measurements on dilute samples, a considerable uncertainty is present in both approaches.[32]

Alternatively, we tested a novel approach developed in our recent survey,[33] which relies on the fact, that HS prefer to sit at contact in the fluid state and shows very little systematic dependence on the degree of polydispersity, as long as $\sigma \leq 0.07$. We therefore determined the static structure factor, $S(q)$, over a volume fraction range $0.34 \leq \Phi \leq 0.45$ as obtained from phase diagram mapping. Next, the positions $q_{MAX}$ of the maximum in $S(q)$ were read off from the data and plotted against the un-swollen volume fractions $\Phi_u$. The curve of $q_{MAX}$ versus $\Phi_u$ was then mapped onto the theoretically expected dependence of $q_{MAX}$ on $\Phi$ calculated by theoretical expressions based on the Verlet–Weis-corrected Percus-Yevick integral equation for polydisperse HS, varying the polydispersity $\sigma$ parametrically between 0.03 and 0.07. From this procedure, rescaled effective interaction radii, $R_{PY}$, and rescaled volume swelling ratios, $Q_{PY}$, were obtained for each $\sigma$-value and then averaged.

For the chosen combination of particles we obtained $R_{PY,S}$ = 129.4±0.6 nm and $R_{PY,L}$= 166.3±0.7 nm, $Q_{PY,S}$ = 5.96±0.05 and $Q_{PY,L}$ = 5.11±0.05. These values are in good agreement with those obtained from Bragg scattering in the crystal phase: $R_{BR,S}$= 127±4 nm, $R_{BR,L}$= 164±3 nm, and close to those from phase diagram mapping neglecting polydispersity: $R_{PD,S}$ 134±4 nm, $R_{PD,L}$= 174±3 nm The corresponding particle size ratios are: $\Gamma_{PY}$ = 0.778±0.006, $\Gamma_{PD}$ = 0.768±0.04 and $\Gamma_{BR}$ = 0.772±0.05 all in good mutual agreement and consistent with $\Gamma_h$= 0.770±0.02 obtained from the much larger hydrodynamic radii $R_{h,S}$= (148±2) nm and $R_{h,L}$ = (193±2) nm. Therefore, an effective size ratio of $\Gamma_{eff}$ = 0.77±0.005 is assumed in the following, which places our samples close to the maximum extension of the LP - F coexistence region. The volume fraction scale was then set by converting weighed-in mass fractions of S and L into $\Phi_{PY}$ using the rescaled $Q_{PY}$

A stock suspension with $x_S$ = 0.76 (number ratio $Z = N_S/N_L$=3.2; volume fraction ratio $\phi_S/\phi_L$=1.5) and $\Phi$ = 0.6 was prepared from weighed-in masses $m_{S/L}$ of S and L particles and $m_{2EN}$ of 2EN from the dry particle powders. After letting the particles swell for several weeks, two series of samples were prepared by dilution respecting $\Phi = [(m_S/\rho_{PS}) \cdot Q_{HS,S} + (m_L/\rho_{PS}) \cdot Q_{HS,L}] / [(m_S+m_L+m_P)/\rho_{PS} + m_{2EN}/\rho_{2EN}]$. Here, $m_P$ denotes the mass of linear PS with of molar mass $M_w$ = 133000 g mol$^{-1}$ and a radius of gyration $R_G$ = 13 nm added to induce attraction. Samples were then put on a tumbling wheel to allow for size equilibration and to keep the suspensions in a homogenized shear molten state prior to the experiments.

Repulsive (R-samples) samples were prepared at volume fractions of 0.552 ≤ $\Phi$ ≤ 0.582. The upper value was chosen to be close to the repulsive glass transition occurring at $\Phi$ ≈ 0.585 as observed in dynamic light scattering. Sample locations are shown in FIG. 1 to start in the LP - F coexistence region, cross the three phase coexistence region LP - F - S (With S denoting fcc crystals of the S component) and enter the LP - S coexistence region. A second sample series was prepared at a colloid volume fraction $\Phi$ = 0.578 and polymer concentrations of (0-16.9) mg/mL (A-Samples).

All static light scattering experiments were performed using a modified commercial instrument (SOFICA, France) described previously in some detail.[30, 33] Here we only note that the angular resulution was 1° over a range of 25-145°, yielding scattering vector ranges of q = (5-30) μm$^{-1}$ and q = (10-50) μm$^{-1}$ for wave lengths of λ=633 nm or λ=403 nm, respectively. Taking the samples off the tumbler defines $t_W$ = 0.

Crystals appeared within weeks but took several months to grow to their final size. In the low-$\Phi$ samples the crystals settled within month to the cell bottom. In all samples, coarsening of crystals was observed and crystals reached sizes up to a mm. For all R-samples we obtained a characteristic scattering pattern with a group of reflections around q ≈ 14 which are identified as super-structure peaks since the main structure factor peak of the initial melt was located at $q_{MAX}$ ≈ 25 µm⁻1. Two examples of late stage diffractogram are shown in FIG. 2.

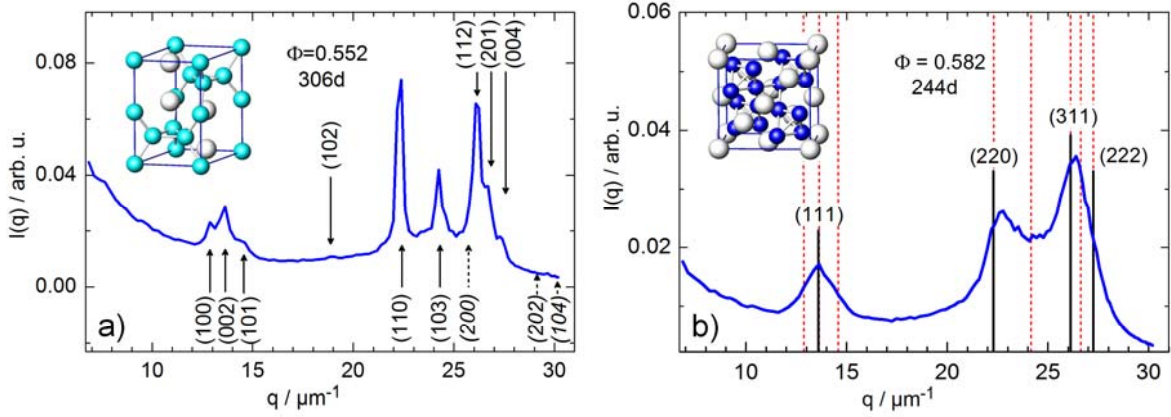

FIG. 2 Late stage powder diffractogramms of two binary HS mixtures with $x_S$ = 0.76. a) R-01 at $\Phi$ = 0.552 and $t_W$ = 306 d. Peak locations as calculated for a hexagonal structure are shown by Miller indexed arrows. Inset: conventional unit cell of MgZn2 with white balls representing Mg and blue balls representing Cu. b) R-7 at $\Phi$ = 0.582 and $t_W$ = 224 d. Peak locations as calculated for a hexagonal structure are shown by vertical red-dotted lines. Locations as calculated for a cubic structure are shown by Miller indexed thick black lines. Inset: conventional unit cell of $MgCu_2$ with white balls representing Mg and blue balls representing Cu.

FIG. 2a shows a sample at $\Phi$ = 0.552 and $t_w$ = 306 d. Assuming a hexagonal structure and assigning the first two peaks at 12.9 µm⁻¹ and 13.6 µm⁻¹ to (100) and (002) results in lattice parameters a=564 nm and c=922 nm, respectively. Using these values, we calculated further peak positions according to:

$$q = \frac{2\pi}{a}\sqrt{\frac{4}{3}(h^2+hk+k^2)+\left(\frac{a}{c}\right)^2 l^2} \qquad (1),$$

and compare these to the observed peaks in FIG. 2a. All observed peaks can be indexed. The (102) reflection at 18.80 µm⁻¹ is weak but clearly visible, (001), (003) and (111) are absent. The (200) reflection (expected at 26.1 µm⁻¹) is hidden in the flank of the strong (112) reflection and the two higher order reflections (202) and (104) are too weak to be identified. From this analysis, we identify the crystallographic space group as $P6_3/mmc$. The axial ratio of c/a = 1.636 is consistent with the ideal axial ratio for the MgZn2 LP of c/a = $(8/3)^{1/2}$=1.633,[48] but not with that of $MgNi_2$. Moreover, none of the diffractogramms shows the (001) reflection. $MgNi_2$ can therefore be excluded. With 8 S and 4 L particles in the unit cell of $MgZn_2$ the crystal packing fraction amounts to $\Phi_C = 0.590$ for radii $R_{PY}$ taken from structure factor analysis and $\Phi_C = 0.575$, for radii $R_{BS}$ from crystal Bragg reflections at coexistence.

FIG. 2b shows a sample at $\Phi = 0.582$ and $t_w = 224$ d. Three broad and featureless main peaks are observed and the overall intensity is smaller than in FIG. 2a. Assigning the first peak at 13.63 µm$^{-1}$ to (002), the red dotted lines give the expected peak positions for a hexagonal structure. While all main peaks can be indexed, the formerly prominent (103) peak is missing. We therefore tried a cubic indexing by taking the first peak to be (111). The resulting peak locations coincide with hexagonal reflections and are marked in FIG. 2b as black lines. Based on this indexing, the observed diffractogram is not incompatible with the crystallographic space group $Fd\,3m$ corresponding to $MgCu_2$, even though the broad maxima of the two higher order reflections do not quantitatively coincide with the expectations. Also for all other R-samples, LPs are present. Recorded diffractograms either resembled the cases shown in FIG. 2 or appeared to be a superposition. This also occured for individual samples as $t_W$ increased. We attribute this kind of switching behaviour to the settling of crystals, which stay in the detection volume only for limited amounts of time. In all samples, however, $MgNi_2$ could be excluded and $MgZn_2$ confirmed, while the additional presence of $MgCu_2$ could not be ruled out. In none of the samples at larger $\Phi$, we could observe pure S-crystals formation.

Very similar diffractogramms could also be observed in the attractive A-samples for small polymer concentrations of $c_P < 2$ mg/mL (FIG. 4d). For larger polymer concentration no crystal formation could be observed in a range 2 mg/mL $\leq c_P <$ 7.44 mg/mL. For 7.44 mg/mL $\leq c_P \leq$ 16.91 mg/mL we again observed crystal formation. The coresponding diffractogramms in FIG. 3, however, contain no super-structure peaks. Instead, the initial broad fluid maximum around $q_{MAX} \approx 22$ µm$^{-1}$ is shifted to values around 18 µm$^{-1}$, and a group of three peaks evolves at scattering vectors q $\approx 30$ µm$^{-1}$. With increasing $c_P$ the time scale of crystallization is reduced, but the peaks get broader loose in intensity until for $c_P = 16.91$ mg/mL only a weak single peak remains. A late stage diffractogramm for sample A-12 at $c_P = 10.9$ mg/mL was measuured at the shorter wavelength at $t_W = 304$ d and is displayed in FIG. 3b. In addition to the triplett, here, the flank of a fourth peak at q $\approx 48$ µm$^{-1}$ is visible, unfortunately just on the border of the observation range. Miller indexing can be applied for a simple hexagonal close packed (hcp) structure formed by the S-component. From the location of the (100) and

(002) reflections we obtain lattice constants a = 260 nm and c = 427 nm yielding an axial ratio c/a = 1.645 and a $R_{PY,S}$-based crystal volume fraction of $\Phi_C$ = 0.73. With these values the remaining reflections can be identified as (101) and (110) reflections, whereas the (102) reflection is missing.

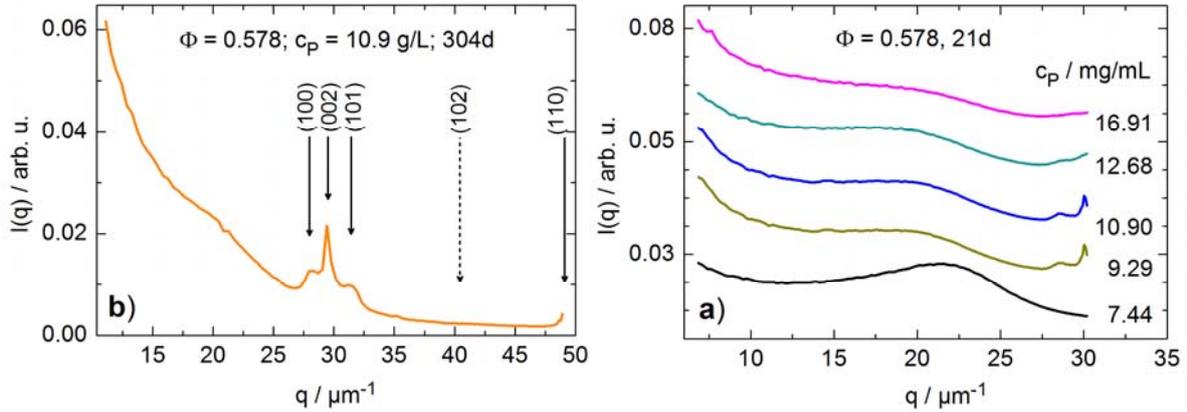

FIG. 3. a) Diffractogramm measured at λ = 403 nm of sample A-12 with Φ = 0.578, $c_P$ = 10.9 mg/mL and $t_W$ = 304 d. Peak locations as calculated for a hexagonal structure are shown by Miller indexed arrows. b) Diffractogramms of different a-samples at Φ = 0.574 and large $c_P$ as indicated. Formation of dense r-hcp crystals is fastet for the samples with $c_P$ = (9-11) mg/mL. The low-$c_P$ sample has not yet started to crystallize. During crystallization a broad, fluid like peak at q-values below $q_{MAX}(t_W = 0)$ evolves in parallel to the high-q group of reflections.

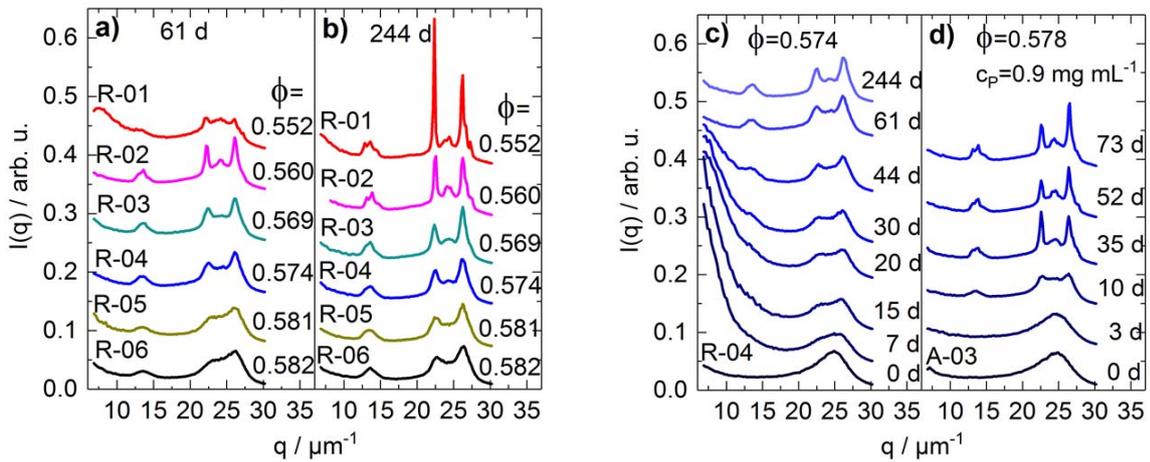

FIG. 4. Comparison of diffractograms. Data are shifted for clarity. a) Different samples of the R-series with volume fractions as indicated at $t_W$ = 61 d. b) The same samples at $t_W$ = 244 d. c) Time series of sample R-04 at Φ = 0.574 and times indicated. d) the same for sample A-03 at Φ = 0.574 and $c_P$ = 0.9 mg/mL. For details see text.

While both the repulsive and the weakly attractive samples form LPs, these two series show strikingly different crystallization kinetics and resulting crystal quality. This is demonstrated in FIG 4 a-d. First, in FIG. a and b, we compare the diffractogramms of different R-samples for two different times. At early times, crystal formation is most progressed for R-02, while for R-01, the low-q superstructure peak as well as (110) and (112) have just become visible. At later times, all peaks have grown in intensity and gained in sharpness for all samples, but only for R-01 and R-02 the peaks are narrow enough to be clearly discriminated. In particular, the pronounced increase in intensity and the corresponding peak narrowing seen for R-01 is striking. In fact, after 244 days the homogeneously nucleated crystals had grown to nearly mm size in this sample. Visual inspection further showed that after crystal settling, no further nucleation events occurred and the final phase equilibration occurred *via* a columnar crystal growth on top of the settled crystals. These columns were found to grow to several mm$^3$ in volume. In FIG. 4c and d, we compare two time series: c) shows an an attractive sample; d) shows a repulsive sample of similar $\Phi$. For R-04, we initially observe a strong contribution of forward scattering, which, however, vanishes over the period 30 d $\leq$ tW $\leq$ 60 d during which the super.-structure peaks appear. For A-03, we observe a much earlier appearance of the typical LP diffractogram which after two months shows a fully developed peak structure. In the attractive sample, a strong forward scattering contribution is absent. Already this preliminary comparison shows that the formation of LPs involves complex and strongly interaction dependent kinetics. A more detailed study is under way.

The two main results of this study can thus be summarized. Buoyant HS-approximants of size ratio $\Gamma = 0.77$ spontaneously form Laves phase crystals in excellent agreement with the expectation from simulations on ideal HS. Addition of non-adsorbing polymer accelerates the crystallization kinetics and improves the crystal quality.

A few points need further attention. First, the very good to near quantitative agreement with the theoretical predictions is remarkable. All three LPs differ only in the stacking sequence of the same four-layered structural units. In the the MgZn$_2$ structure, this packing is AABB, in the MgCu$_2$ structure AABBCC, and in the MgNi$_2$ structure in Fig. AABBAACC. They have the same maximum packing fraction of $\Phi_{MAX} = 0.73$ at $\Gamma_{MAX} = 0.818$.[49] The decrease of $\Phi_{MAX}$ with decreasing $\Gamma$ is slightly less pronounced for MgZn$_2$ than for MgCu$_2$ but significantly more pronounced for MgNi$_2$. The decrease for $\Gamma > \Gamma_{MAX}$ is the same for all three LPs. On the high-$\Gamma$ side, the three phases differ only marginally with respect to thermodynamic stability. For $\Gamma = 0.82$ and $\Phi = \phi_S + \phi_L = 0.6$ MgZn$_2$ has the lowest bulk free energy per particle at 7.436 k$_B$T, followed by MgNi$_2$ at 7.438 k$_B$T, and MgCu$_2$ at 7.439 k$_B$T.[6, 10] The authors therefore expected that experiments should show a mixture of all three LPs similar to the ex-perimental observation of the close-packed randomly stacked r-hcp crystals of pure hard spheres, which can be seen as a mixture of fcc and hcp crystals. Concerning the formation of single component r-hcp phases in the presence of non-adsorbing polymers with size ratio $\xi = R/R_g = 0.1$, theory and simulation predict lowered transition pressure[52] and a compression to the HS packing limit of $\Phi_{HCP,MAX} = 0.74$.[50, 51]

In the present study, we can clearly identify the MgZn$_2$ structure for the repulsive samples crystallizing in the F-LP coexistence region at low $\Phi$, as well as for weakly attractive samples at $\Phi = 0.578$. and exclude the MgNi$_2$ structure. Diffractograms of repulsive samples at large $\Phi$ appear to be compatible with the presence of MgCu$_2$. We worked at $\Gamma = 0.77 < \Gamma_{MAX}$, where the maximum theoretical packing fractions are about 0.67 for MgZn$_2$ and MgCu$_2$ and about 0.65 for MgNi$_2$. The packing fraction of the experimental LPs in the F-LP coexistence region was very close to 0.59, i.e. the packing fraction of the LP-corner of the F-LP coexistence triplet in FIG. 1b. The presence of MgZn$_2$ at low $\Phi$ suggests the lowest free energy for this structure also for $\Gamma < \Gamma_{MAX}$. The absence of MgNi$_2$ suggests a much larger free energy, compatible with the calculations of the maximum packing fraction at $\Gamma = 0.77$. Our finding of MgCu$_2$ compatible diffractogramms at large $\Phi$ and mixed forms in between supports the suggestion of Hynninen et al. that randomly stacked LPs can be formed. Since the experiments were done at $\Gamma < \Gamma_{MAX}$, however, these are restricted to randomly alternating MgZn$_2$ and MgCu$_2$. Concerning the compression of the small component in the L-S eutectic region we here observe a value of $\Phi_C = 0.73$ again in very good agreement with the predictions. The small deviation is attributed to polydispersity induced perturbations of the ideal packing.

Second, our study explored the phase behaviour of binary mixtures of attractive HS, in regions not yet covered theory. Our findings are compatible with the phase behaviour expected for repulsive mixtures at increased pressures. A F-LP coexistence region is followed by a non-crystallizing region and a region, where pure S forms. This is clearly compatible with the phase sequence depicted in FIG 1a, if the intermediate region is identified with LP-S coexistence. Here two incompatible crystal structures compete, i. e. while LPs afford positional fluctuations for nucleus formation, S-hcp affords additional composition fluctuations. Both high pressure (high $c_P$) phases are not reached in the repulsive samples, where for this mixture, a kinetic glass transition intervenes at $\Phi = 0.585$. Our observation therefore suggest, that depletion attractive HS mixtures should show very similar phase behavior as their repulsive counterparts, but all phase boundaries will appear shifted towards lower $\Phi$ in the $\phi_L$ - $\phi_S$ diagram and to lower P in the P-$x_S$ plane. This conclusion seems to be further supported by our previous investigation[30] of a binary attractive HS mixture with $\Gamma$ = 0.74 and by the observation of lowered freezing pressures in computer simulations on single component HS and attractive HS systems.[52]

Addition of polymer not only shifts the phase boundaries. Our study also shows that the crystallization kinetics is favorably affected. Previous studies reported an increased mobility in attractive systems to induce a reentrant glass transition with a highly mobile fluid at intermediate and an attractive glass at large $c_P$.[53, 54] Further, attraction is theoretically expected to assist fractionated crystallization of single component systems which occurs for large polydispersity $\sigma \gtrsim 0.07$,[55] and for which a sorting by size is required. This effect has been observed experimentally for the individual polydisperse components of a eutectic binary mixture.[56] Moreover, sorting into L and S was also favored in the eutectic region of the same mixture.[30]

In the experiments, we observe the crystallization of the pure S-component from the mixed melt at 7.44 mg/mL ≤ $c_P$ ≤ 12,68 mg/mL. At $c_P$ = 10,90 mg/mL, differentiated crystal reflections are clearly visible already after 21 d, while the corresponding R-sample shows a (LP-) peak differentiation only after some two months. Therefore, also here, crystallization involving a sorting by size is accelerated, if only a single crystal structure (r-hcp) is possible for the separated component. On the other side, the accelerated nucleation and growth of non-close-packed LPs at low $\Phi$ and weak attraction shows that attraction can also speed up local ordering. An interesting situation therefore emerges, where simultaneous sorting and local ordering compete. This is the case in the range of $c_P$ between the regions of LP-F and S-L coexistence. There, a coexistence of LP and S can be expected from the repulsive phase diagrams. Thus two crystal structures simultaneously compete for S-particles. However, the complex influence of depletion attractions on the crystallization kinetics certainly needs further experimental and theoretical attention.

Finally, it may be interesting to note that this is the first report on HS-LPs, despite their great potential for the fabrication of photonic materials and intense experimental and theoretical investigation. Unlike charged spheres,[20, 21, 22] where the particles stay at considerable mutual distances and unlike nano-particle systems,[25, 26, 27, 28] where the lattice constants are much smaller than desired in photonics, HS-LPs can form at contact and with lattice constants comparable to the wave length of visible light, which simplifies further processing.[6] Such requirements have so far only been met by natural gem opals, for which, however, the synthesis conditions are unexplored.[24] Their realization in the present work on one side relied on the simultaneous availability of precise and sufficiently extensive theoretical predictions which allowed identification of the most promising size ratio, molar fraction and colloid-polymer size ratio for a successful investigation.[9, 49, 51] On the other side, our particles were chosen after extensive characterization and prepared as close as possible to the boundary conditions of the simulations. I. e. we employed excellently buoyancy matched hard sphere approximants with steeply repulsive pair potentials. Moreover, we employed our recently introduced approach,[33] in which we determined the effective particle radii $R_{PY}$ via static structure factors measured in the fluid phase of the pure components. This allowed choosing the size ratio free of any bias by component polydispersity or surface roughness. As a result all investigated systems spontaneously formed the expected LPs and the determined coexistence volume fractions showed a near quantitative agreement with the theoretical expectations for the respective LP and S crystals. Certainly, this cannot be reached for size ratios determined by other precise, yet inaccurate procedures. A case in point is our previous study,[30] which had aimed at a size ratio of 0.785 but found a eutectic behavior, which in fact is expected for the corrected $R_{PY}$-based size ratio of $\Gamma = 0.74$.[10]

Concluding, we have combined existing theoretical results with to-the-point preparation of binary HS mixtures to grow large Laves phase crystals of $MgZn_2$ structure. We employed near µ-g conditions and worked in the F-LP coexistence to remove possible kinetic obstacles to crystallization like the kinetic glass transition or competing crystallization of other structures. Addition of polymer was observed to retain the phase behavior but moreover to speed up crystal formation and improve crystal quality. We therefore anticipate that our approach is of great interest for an improved understanding of crystallization kinetics in classical systems and useful for the realization of photonic materials.

Acknowledgements

We thank D. Sagawe, J. Schneider, M. Wiemann, and A. Rabe for their assistance in particle synthesis as well as M Röhr and R. Beyer for fruitful discussions on the interpretation of the diffractogramms. Financial support by the DFG (Grants Nos. Ba1619/2 aand Pa459/13) is gratefully acknowledged.

References

1   F. Laves, *Naturwissenschaften*, 1939, **27**, 65-73.

2   R. L. Berry and G. V Raynor, *Acta Cryst.*, 1953, **6**, 178-186.

3   R. L. Johnston and R. Hoffmann, *Z. Anorg. Allg. Chem.,* 1992, **616**, 105-120.

4   F. Stein, M. Palm and G. Sauthoff, *Intermetallics*, 2004, **12**, 713-720.

5   F. Stein, M. Palm and G. Sauthoff, *Intermetallics*, 2005, **13**, 1056-1074.

6   A.-P. Hynninen, J. H. J. Thijssen, E. C. M. Vermolen, M. Dijkstra and A. van Blaaderen, *Nature Mater.*, 2007, **6**, 202-205.

7   W. G. Hoover and F. H. Ree, *J. Chem. Phys.* **49**, 3609-3617 (1968).

8   M. D. Eldridge, P. A. Madden and D. Frenkel, *Nature*, 1993, **365**, 35-37.

9   M. Dijkstra, in S. Rice, A. R. Dinner (Eds.), *Adv. Chem. Phys.*, 2015, **156**, 35-71.

10  A.-P. Hynninen, L. Filion and M. Dijkstra, *J. Chem. Phys.*, 2009, **131**, 64902.

11  P. Bartlett, *J. Phys.: Condens. Matter*, 1990, **2**, 4979-4989.

12  X. Cottin and P. A. Monson, *J. Chem. Phys.*, 1997, **107**, 6855–6858.

13  S. Punnathanam and P. A. Monson, *J. Chem. Phys.*, 2006, **125**, 24508.

14  S. Hachisu, *Phase Transitions*, 1990, **21**, 243-249.

15  P. Bartlett, R. H. Ottewill and P. N. Pusey, *Phys. Rev. Lett.,* 1992, **68**, 3801-3805.

16  A. Meller, J. Stavans, *Phys. Rev. Lett.,* 1992, **68**, 3646-3649.

17  S. M. Underwood, W. van Megen and P. N. Pusey, *Physica A,* 1995, **221**, 438–444.

18  N. Hunt, R. Jardine and P. Bartlett, *Phys. Rev. E*, 2000, **62**, 900–913.

19  N. Vogel, M. Retsch, C.-A. Fustin, A. del Campo and U. Jonas, *Chem. Rev.*, 2015, **115**, 6265-6311.

20  S. Yoshimura and S. Hachisu, *Prog. Colloid Polym. Sci.*, 1983, **68**, 59–70.

21  M. Hasaka, H. Nakashima and K. Oki, *Trans. Japan Inst. Met.*, 1984, **25**, 65-72.

22  B. Cabane, J. Li, F. Artzner, R. Botet, C. Labbez, G. Bareigts, M. Sztucki and L. Goehring, *Phys. Rev. Lett.*, 2015, **116**, 208001.

23  G. H. Ma, T. Fukutomi and N. Morone, *J. Colloid Interface Sci.*, 1994, **168**, 393–401.

24  J. P. Gauthier, E. Fritsch, B. Aguilar-Reyes, A. Barreau and B. Lasnier, *Comptes Rendus - Geosci.*, 2004, **336**, 187–196.

25  E. V Shevchenko, D. V Talapin, N. a Kotov, S. O'Brien and C. B. Murray, *Nature*, 2006, **439**, 55–59.

26  E. V Shevchenko, D. V Talapin, C. B. Murray and S. O'Brien, *J. Am. Chem. Soc.*, 2006, **128**, 3620–3637.

27  Z. Chen and S. O'Brien, *ACS Nano*, 2008, **2**, 1219-1229

28  W. H. Evers, B. De Nijs, L. Filion, S. Castillo, M. Dijkstra and D. Vanmaekelbergh, *Nano Lett.*, 2010, **10**, 4235–4241.

29  C. P. Royall, W. C. K. Poon, and E. R. Weeks, 2013, *Soft Matter*, **9**, 17-27.

30  A. Kozina, D. Sagawe, P. Díaz-Leyva, E. Bartsch and T. Palberg, *Soft Matter*, 2012, **8**, 627

31  A. Kozina, D. Sagawe, P. Díaz-Leyva, E. Bartsch and T. Palberg, Correction submitted to *Soft Matter*, ArXive submission on 2017/02/18, final number assignement pending

32  W. C. K. Poon E. R. Weeks, C. P. Royall, *Soft Matter*, 2012, **8** 21-30.

33   J. Schneider, M. Wiemann, A. Rabe and E. Bartsch, *Soft Matter*, 2017, **13**, 445–457.

34   S. Golde, T. Palberg, H. J. Schöpe, *Nature Physics*, 2016, **12**, 712–717.

35   H. Senff and W. Richtering, *J. Chem. Phys.*, 1999, **111**, 1705–1711.

36   E. Lange, J. B. Caballero, A. M. Puertas and M. Fuchs, *J. Chem. Phys.*, 2009, **130**, 174903.

37   A. Le Grand and G. Petekidis, *Rheol. Acta*, 2008, **47**, 579–590.

38   P. N. Pusey, W. van Megen, *Nature,* 1986, **320**, 340-342.

39   S. E. Paulin, B. J. Ackerson and M. S. Wolfe, *J. Colloid Interface Sci.*, 1996, **178**, 251.

40   J. L. Harland and W. van Megen, *Phys. Rev. E*, 1997, **55**, 3054–3067.

41   P. Bolhuis, D. Kofke, *Phys. Rev. E*, 1996, **54**, 1.

42   P. Bartlett, *J. Phys.: Condens Matter*, 2000, **12**, A275.

43   M. Fasolo, P. Sollich, *Phys. Rev. Lett.*, 2003, **91**, 068301.

44   P. Sollich, N. Wilding, *Phys. Rev. Lett.*, 2010 **104**, 118302.

45   C. Urban, P. Schurtenberger, *J. Colloid Interface Sci.*, 1998, **20**, 150-158.

46   M. Fasolo and P. Sollich, *Phys. Rev. E*, 2004, **70**, 041410.

47   N. Schaertl, M. Wernet, T. Palberg and E. Bartsch, in preparation

48   J. B. Friauf, *Phys. Rev.*, 1924, **29**, 34–40.

49   L. Filion and M. Dijkstra, *Phys. Rev. E*, 2009, **79**, 46714

50   H. N. W. Lekkerkerker, W. C. K. Poon, P. N. Pusey, A. Stroobants, P. B. Warren, *Europhys. Lett.*, 1992, **20**, 559-564.

51   M. Dijkstra, R. van Roij, R. Roth, A. Fortini, *Phys. Rev. E*, 2006, **73**, 041404.

52   T. Zykova-Timan, J. Horbach, and K. Binder, *J. Chem. Phys.,* 2010, **133**, 014705.

53   T. Eckert , E. Bartsch, *Phys. Rev. Lett.*, 2002, **89**, 125701.

54   K. N. Pham, A. M. Puertas, J. Bergenholtz, S. U. Egelhaaf, A. Moussaïd, P. N. Pusey, A. B. Schofield, M. E. Cates, M. Fuchs, W. C. K. Poon, *Science*, 2002, **296**, 104-106.

55   M. Fasolo and P. Sollich, *J. Chem. Phys.*, 2005, **122**, 074904.

56   A. Kozina, P. Diaz-Leyva, E. Bartsch, T. Palberg, *Soft Matter*, 2014, **10**, 9523-9533.